\documentclass[times,1p]{elsarticle}
\usepackage{amssymb, amsmath, amsbsy, wasysym}
\usepackage[T1]{fontenc}
\usepackage{color}
\usepackage{multirow}
\usepackage{longtable}
\usepackage{url}
\usepackage{algorithmic}
\usepackage[algo2e,linesnumbered,ruled,lined,boxed]{algorithm2e} 

\journal{Journal of Computational Physics}

\newcommand{\algLBM}{%
  \begin{minipage}{.49\linewidth}
  \begingroup
\begin{small}
    \begin{algorithm2e*}[H]

	\SetKwBlock{StreamAndCollide}{Kernel stream and collide}{}
	\SetKwFor{ParFor}{for}{do in parallel}{}
	\SetKw{Gather}{gather}
	\SetKw{Compute}{compute}
	\SetKw{Store}{store}
	\SetKw{Swap}{swap}

		\ParFor{node locations $\vec{x}$}
		{
				\For {all directions $k$}
				{
					\Gather  $\  f'_k \  \leftarrow \  f_k^{pre} (\vec{x} - \vec{c_k})$
				}
				
				\Compute $\rho$ and $\vec{u}$ from $f'_k$

				\For {all directions $k$}
				{
					\Compute $f^{eq}_k$ from $\rho, \vec{u}$

					\Compute $f''_k$ from $f'_k, f^{eq}_k$

					\Store  $\  f^{post}_k (\vec{x}) \  \leftarrow \  f''_k$
				}
		}

		\ParFor{node locations $\vec{x}$}
		{
			\Swap $\  f^{post}_k (\vec{x}) \  \rightleftarrows \  f^{pre}_k (\vec{x})$
		}

	\caption{Standard LBM.}%
\label{alg_LBM}%
\end{algorithm2e*}%
\end{small}
\endgroup
 \end{minipage}%
}

\newcommand{\cmdalgLBMTau}{%
  \begin{minipage}{.50\linewidth}
  \begingroup
\begin{small}
    \begin{algorithm2e*}[H]

	\SetKwBlock{StreamAndCollide}{Kernel stream and collide}{}
	\SetKwFor{ParFor}{for}{do in parallel}{}
	\SetKw{Gather}{gather}
	\SetKw{Compute}{compute}
	\SetKw{Store}{store}
	\SetKw{Swap}{swap}

		\ParFor{node locations $\vec{x}$}
		{
				\For {all directions $k$}
				{
					{\color {black} \Gather  $\  \vec{u}_k \  \leftarrow \  \vec{u}^{pre} (\vec{x} - \vec{c_k})$}
					{\color {black} \Gather  $\  \rho_k    \  \leftarrow \  \rho^{pre} (\vec{x} - \vec{c_k})$}

					{\color {black} \Compute $f^{eq}_k$ from $\rho_k, \vec{u}_k$}

                    $\rho'$ +=  $f^{eq}_k$

                    $\vec{u}'$ += $f^{eq}_k \cdot \vec{c_k}$                    

				}
				
				{\color {black} 	\Store $\rho^{post} (\vec{x}), \vec{u}^{post}(\vec{x}) \  \leftarrow \  \rho', \vec{u}'$ }
		}

		\ParFor{node locations $\vec{x}$}
		{
			{\Swap $\  \vec{u}^{post}(\vec{x}) \  \rightleftarrows \  \vec{u}^{pre}(\vec{x})$} $\ \qquad $
			{\Swap $\  \rho^{post} (\vec{x})   \  \rightleftarrows \  \rho^{pre} (\vec{x})$}
		}

  \caption{%
	Tau1.
}%
\label{alg_LBMTau1}%
\end{algorithm2e*}%
\end{small}
\endgroup
\vspace*{\fill}
 \end{minipage}%
}

\begin{document}
\title{The lattice Boltzmann method for fluid flows at relaxation time equal one: performance study}

\author{{Maja Bacza}}
\ead{majabacza@gmail.com}
\author[pwr]{Tadeusz Tomczak}
\ead{tadeusz.tomczak@pwr.edu.pl}
\author[ift]{Maciej Matyka\corref{cor}}
\ead{maciej.matyka@uwr.edu.pl}
\cortext[cor]{Corresponding author}

\address[ift]{Faculty of Physics and Astronomy, University of Wroc{\l}aw, pl.\ M.\ Borna 9, 50-205 Wroc{\l}aw, Poland, tel.: +48713759357, fax: +48713217682}
\address[pwr]{Wrocław University of Science and Technology, Wybrzeże Wyspiańskiego 27, 50-370 Wrocław, Poland}

\begin{abstract}
Running large-scale computer codes for huge fluid flow problems requires not only large supercomputers but also efficient and well-optimized computer codes that save the resources of those supercomputers.
This paper evaluates the high-performance implementation of the recently proposed Lattice Boltzmann Method (LBM) algorithm with a fixed viscosity and relaxation time $\tau=1$ called Tau1. We show that the performance of the Tau1 algorithm is almost $4\times$ higher than other state-of-the-art standard LBM implementations. We support this finding by detailed complexity analysis and performance study based on the code with several optimizations, including multithreading, vector processing, significantly decreased number of divisions, and dedicated memory layout. We studied its performance in porous media flow in the three-dimensional model of porosity-varying porous medium to make it sound in the physical context. We find performance drops with porosity, which we link to the fact that memory access patterns change dramatically with the increased complexity of the pore space. In contrast to standard LBM implementations, where the performance drops with the number of lattice links, 
the processing speed of the new algorithm measured in fluid node updates per second is almost constant regardless of lattice arrangements.
\end{abstract}


\maketitle

\section{Introduction}
\label{sec1}

Many real-life fluid flow problems require large computational grids, long simulation runs, and highly optimized codes to be efficiently run on supercomputers \cite{Gmeiner2015,Schornbaum2016}. The lattice Boltzmann method (LBM) is a numerical method used to study fluid flow problems, especially useful in complex geometries such as those found in porous media \cite{Succi01,Aidun10} and artery system \cite{djukic2023validation}. The method uses discrete velocity particle distribution, which is evaluated according to the discrete form of the Boltzmann equation \cite{Chen92,Qian92,Aidun10}. In addition, it proved its accuracy in simulating fluid flows in various conditions, including turbulence and complex flows \cite{Chen92,Qian92,Succi01}. Although the computational complexity of the LBM algorithm is relatively high \cite{Tessarotto2008}, we can parallelize it to achieve high performance. 

The recently investigated lattice Boltzmann method that simplifies the original algorithm by assuming a unit relaxation time $\tau\!=\!1$ does not require the storage of the distribution function, making it easier to implement and requiring less memory \cite{tau1}. Because of fixed viscosity, however, the Reynolds number may only be controlled by the characteristic velocity and length scale. Thus, it is necessary to enlarge grids for higher Reynolds numbers, and this may limit its ability to accurately simulate high-velocity flows, for example, where inertial effects occur  \cite{Andrade99}. 
Due to the larger grids, the algorithm may require more operations per iteration and more data access operations, which could result in a significant performance drop \cite{tau1}.
Thus, the Tau1 method must be optimized to improve its accuracy and efficiency to make it practically valuable for large systems.
Our work aims to design and evaluate the high-performance Tau1 implementation that could solve problems efficiently at large grids.
We use standard optimization techniques, including Single Instruction Multiple Data (SIMD) processing and a simple memory layout mitigating AMD Zen 4 architecture performance penalties. We use the Zou-He pressure boundary conditions and halfway boundary conditions on the walls and store velocities instead of momentum to limit the number of division operations.
We test the algorithm and fully optimized code in standard 2D benchmark cases and 3D porous media samples to test if optimization procedures did not affect the accuracy of the solution. Our study shows an increased algorithm performance at $\tau=1$ compared to the standard LBM in simple benchmark flows. We further tested this in porous media and showed a performance drop with decreasing porosity. The overall conclusion is that by properly optimizing the Tau1 algorithm, we can achieve an algorithm that is almost four times faster than the standard LBM formulation.

\section{The Lattice Boltzmann Method}
\label{LBMdescription}

The Lattice Boltzmann (LBM) algorithm extends lattice gas automata. It uses continuous particle distributions \cite{Succi01, Qian92} to model fluids. It works on a grid and comprises two main steps: propagation and collision. The LBM uses spatial, velocity, and time discretization. The velocity discretization uses a set of velocity directions. For example, D2Q9 (two-dimensional set of nine velocity directions) may be used in 2D and D3Q19 or D3Q27 for the 3D case. Continuous particle distributions stay in the regular grid, with each distribution related to the population of particles moving in a given velocity direction. For example, in an $L\times L$ grid using the nine velocity D2Q9 scheme, at least $9L^2$ particle distributions are kept in memory.

The Lattice Boltzmann equation, which discretizes the Boltzmann equation over time, space, and velocity \cite{He97}, leads to the LBM Method defined by the equation:
\begin{equation}
f_i(\vec{x}+\vec{c_i}\Delta t, t + \Delta t) = f_i(\vec{x}, t) + \Omega_i(\vec{x}, t),
\label{LBM_equation}
\end{equation}
where $f_i$ is the distribution function, $\vec{x}$ is the position, $\vec{c_i}$ is the velocity, and $\Omega_i$ is the collision operator. In its most common form, the LBM uses the single relaxation time Bhatnagar-Gross-Krook (BGK) operator given by:
\begin{equation}
\Omega(f) = \frac{f_{eq} - f}{\tau},
\label{BGKoperator}
\end{equation}
where $f_{eq}$ is the equilibrium distribution function and $\tau$ is the relaxation time. The equilibrium distribution function describes the local equilibrium state towards which the system evolves due to collisions. The relaxation time controls the viscosity of the modeled fluid.
The equilibrium function is given by:
\begin{equation}
f_i^{eq}(\vec{x}, t) = w_i\rho\left(1+\frac{\vec{u}\cdot\vec{c_i}}{c_s^2}+\frac{(\vec{u}\cdot\vec{c_i})^2}{2c_s^4}-\frac{\vec{u}\cdot\vec{u}}{2c_s^2}\right),
\label{feq}
\end{equation}
where $w_i$ are model-specific weights, dependent on the velocity set ${\vec{c_i}}$, $c_s$ is the speed of sound ($c_s=1/3\frac{\triangle x^2}{\triangle t^2}$), $\vec{u}$ is the local velocity, and $\rho$ is the local density.
The BGK operator given in Eq.~(\ref{BGKoperator}) is substituted into Eq.~(\ref{LBM_equation}) with a time step of $\Delta t = 1$, resulting in
    \begin{equation}
      f_i(\vec{x}+\vec{c_i}, t + 1) = f_i(\vec{x}, t) + \frac{f_i^{eq}(\vec{x}, t) - f_i(\vec{x}, t)}{\tau},
    \label{LBGK}
    \end{equation}
which can be divided into two parts: collision and propagation. This separation simplifies the implementation because the first step becomes entirely local.
The macroscopic variables are obtained using the moments of the distribution function. The local density and velocity are given as the sums of the distribution along all velocity directions:
\begin{equation}
        \rho(\vec{x}, t) = \sum_i f_i(\vec{x}, t).
        \label{lbm_density}
\end{equation}
\begin{equation}
        \vec{u}(\vec{x}, t) = \frac{1}{\rho(\vec{x}, t)} \sum_i \vec{c_i} f_i(\vec{x}, t).
        \label{lbm_velocity}
\end{equation}
The pressure is related to the density using the linear equation of state:
\begin{equation}
      p=c_s^2\rho.
\end{equation}
\subsection{The Tau1 algorithm}
\label{tau1_description}
Tau1 is a variation of the LBM method, where the relaxation time $\tau$ is set to one \cite{tau1}.
This assumption leads to a significant simplification of the original transport equation Eq.~(\ref{LBGK}), which, after a few simple algebraic operations, results in
        \begin{equation}\label{LBMTau1}
            \left.f_i(\vec{x}+\vec{c_i}, t + 1)\right|_{\tau=1}  = f_i^{eq}(\vec{x}, t).
        \end{equation}
This procedure removes the dependency of the main transport equation on the previous time step, thereby saving computer memory and bandwidth. Because the equilibrium distribution function depends only on the local velocity and density, the Tau1 solver is similar to the standard computational fluid dynamics (CFD) methods, which, in principle, operate only on primitive variables. 
To obtain the macroscopic variables in Tau1, we compute the in-place sums using the equilibrium function given by Eq.~(\ref{feq}). Thus, for density, we have:
        \begin{equation}
                \rho(\vec{x}, t)|_{\tau=1} = \sum_i f_{inv(i)}^{eq}(\vec{x}+\vec{c}_i, t),
                \label{lbmtau1_density}
         \end{equation} 
and for velocity:
    \begin{equation}
    \vec{u}(\vec{x}, t)|_{\tau=1} = \sum_i f_{inv(i)}^{eq}(\vec{x}+\vec{c}_i, t)\cdot \vec{e}_k,
    \label{lbmtau1_velocity}
    \end{equation} 
where $inv(i)$ denotes the direction opposite the i-th velocity vector.

\begin{algorithm2e}[t]    
                \DontPrintSemicolon
                \ForAll(\tcp*[h]{$\vec{x}$ are fluid node locations}){$\vec{x}$}{$\vec{u}(t, \vec{x}) \gets (0, 0)$\tcp*[l]{mass flux at $\vec{x}$ at time $t$}
                    $\rho(t, \vec{x}) \gets 0$\tcp*[l]{density at $\vec{x}$ at time $t$}
                    \ForAll(\tcp*[h]{$k$ is velocity index}){k}{
                        \eIf(\tcp*[h]{$\vec{c_k}$=$k$-th velocity}){$\vec{x}-\vec{c_k}$ is a wall}{
                            $ik \gets i\;\;:\vec{c_i}=-\vec{c_k}$\;
                             $f_{ik}^{eq} \gets w_{ik}\rho\left(1+\frac{\vec{u}^*\cdot\vec{c_{ik}}}{c_s^2}+\frac{(\vec{u}^*\cdot\vec{c_{ik}})^2}{2c_s^4}-\frac{\vec{u}^*\cdot\vec{u}^*}{2c_s^2}\right)$\tcp*[l]{$w_k$ is the weight}
                        }
                        {
            {...\tcp*[l]{fluid node}}
            }
                        
                        $\vec{u}(t, \vec{x}) \gets \vec{u}(t, \vec{x}) + e_{ik} \cdot f^{eq}_{ik}$\;
                        $\rho(t, \vec{x}) \gets \rho(t, \vec{x}) + f^{eq}_{ik}$\;
                    }
                }
                \caption{Complete Tau1 algorithm with halfway bounce-back second-order wall boundary condition.}
                \label{lbmtau1_bounceback}          
            \end{algorithm2e}

\subsubsection{Wall boundary condition}

Using the standard second-order accuracy on the wall in the halfway bounce-back scheme, as previously done in standard LBM \cite{Zou97}, is impossible in the Tau1 algorithm as it stores the values of macroscopic velocity and density fields instead of the distribution function from the previous time step.

Thus, we suggest recomputing the unavailable components of $f_i$ at the wall using Eq.~(\ref{feq}) and implementing the halfway boundary condition from the reconstructed distribution. The reconstructed distribution for a boundary node at $\vec{x}$, based on the available values, will be given by:
\begin{equation}
f_{ik}^{eq} \gets w_{ik}\rho^*\left(1+\frac{\vec{u}^*\cdot\vec{c_{ik}}}{c_s^2}+\frac{(\vec{u}^*\cdot\vec{c_{ik}})^2}{2c_s^4}-\frac{\vec{u}^*\cdot\vec{u}^*}{2c_s^2}\right)
\end{equation}
where $\vec{u}^* = \vec{u}(t-1, \vec{x})$, $\rho^* = \rho (t-1, \vec{x})$ are values of velocity and density at a previous time step, respectively, which replaces line 7 of the original algorithm given in \cite{tau1}. We present the complete Tau1 algorithm with the halfway bounce-back implementation in the Algorithm~\ref{lbmtau1_bounceback}.

\subsubsection{Algorithmic differences between LBM and Tau1}

\begin{figure*}[!ht]
\algLBM \hfill \cmdalgLBMTau
\caption{%
Comparison of the \emph{pull} version of the standard LBM implementation with the AB memory layout (Algorithm \ref{alg_LBM}) with the Tau1 (Algorithm \ref{alg_LBMTau1}).%
}%
    \label{alg_comparison}
\end{figure*}
In order to spot the similarities and differences between algorithms, we first compare the classical \emph{pull} version of the LBM algorithm with two copies of the distribution functions $f_i$ stored in memory (the so-called AB memory layout) to the Tau1 formulation with two copies of the density and velocity values stored in memory \cite{tau1} (see Fig.~\ref{alg_comparison}). In both algorithms, previous versions of read values (the read-only ones) are denoted by the \emph{pre} upper index, and the new values that are currently computed and stored in memory are denoted by the \emph{post} index. Additionally, the notation shown in Fig.~\ref{alg_comparison} mainly focuses on low-level machine operations (memory reads and stores) with computations shown in minimal.

As shown in Fig. \ref{alg_comparison}, both algorithms have similarities. During computations of a single time step, the nodes can be processed independently, making the algorithms embarrassingly parallel. The processing of a single node can be implemented with three main stages: 1) loading the data for neighboring nodes from the external memory (the gather operation), 2) computing the new values from the gathered data, and 3) storing the new values computed for the current node in memory. The complexity of the computations is also similar; in both algorithms, the values of equilibrium functions are computed from density and velocity (line 7 in Algorithm \ref{alg_LBM} and line 4 in Algorithm \ref{alg_LBMTau1}) and the density and velocity are computed from the distribution functions (line 5 in Algorithm \ref{alg_LBM} and lines 5 and 6 in Algorithm \ref{alg_LBMTau1}).

Despite these similarities, there are significant differences. First, in Tau1, more data is gathered from the neighboring nodes. The standard LBM requires only a single distribution function from each of the neighbors, whereas the Tau1 version loads all values of the density $\rho$ and the velocity vector $\vec{u}$ for each neighbor node. However, in the Tau1 formulation, all neighbor nodes of a given node load the same density and velocity values from this node. Thus, it is possible to significantly minimize the overall memory traffic by buffering, inside the processor, the velocities and densities for neighboring nodes (as shown in Section \ref{sec:complexity}). Such data buffering is impossible in standard LBM because each distribution function is used only once per step.

Second, in standard LBM, all gathered distribution functions $f'_k$ must be stored inside the kernel because these functions are used twice in lines 5 and 8 of Algorithm  \ref{alg_LBM}. Thus, the local variables must contain the computed $\rho, \vec{u}$ values and the gathered functions $f'_k$ at some point in the kernel. In Tau1, the gathered values of $\rho_k, \vec{u}_k$ and the computed $f^{eq}_k$ function can be dropped immediately after accumulation in lines 5 and 6 of Algorithm \ref{alg_LBMTau1} limiting the local data stored inside the kernel to values of $\rho'$ and $\vec{u}'$ only. Another advantage of the Tau1 algorithm is that it contains only one internal loop over $k$, which can be parallelized (assuming parallel reduction algorithms in lines 5 and 6 of Algorithm \ref{alg_LBMTau1}). In contrast, standard LBM uses two loops over $k$, and the second loop (lines 6--10 from Algorithm \ref{alg_LBM}) requires a preceding code to compute $\rho$ and $\vec{u}$ (line 5).

Third, the Tau1 algorithm does not contain a relaxation step for the computation of the new distribution functions (line 8 Algorithm \ref{alg_LBM}), which may, in theory, allow saving a few arithmetic operations per node. 

\subsubsection{Complexity analysis}
\label{sec:complexity}

In LBM, D$d$Q$q$ denotes the velocity discretization scheme where $d$ is the space dimension, and $q$ is the number of lattice links. Assuming that the AB memory layout in time is used, with one "old" population and one "new" population, in the Tau1 algorithm, we need to store two copies of both the local density $\rho$ and the local velocity vector $\vec{u}$ (with $d$ coordinates). Note that we ignore additional data needed to store types of nodes (fluid, solid, boundary) owing to a low impact on total memory and bandwidth usage because a node type can be coded only in a few bits per node. The memory required by the Tau1 algorithm is $2 \cdot (d+1) \cdot s$ bytes per lattice node, where $s$ denotes the number of bytes required to store a single value in memory ($s=8$ and $s=4$ bytes for double and float precision, respectively).

The computational complexity of the Tau1 algorithm is difficult to define because, in practice, the arithmetic operations resulting from Eq.~(\ref{feq}) are significantly reorganized during code compilation, particularly with a high level of compiler optimization. Nevertheless, we may estimate the number of required arithmetic operations similar to the standard LBM implementations because, in both cases, we must compute and accumulate the same number of equilibrium functions $f^{eq}_i$. 

\begin{table*}[t]
    \centering
    \caption
    {
        Comparison of theoretical memory and bandwidth usage per node for different LBM implementations.
				AA and AB denote memory schemes (single and two copies of data, respectively), and $s$ denotes the number of bytes per single data value.
        Memory usage and traffic reductions were computed as the ratio of the LBM parameter to the Tau1 parameter.
    \label{tab_complexity}}
     \begin{tabular}{c c c c c c c c c c c }
        \multirow{3}{*}{Lattice} & \multicolumn{3}{c}{LBM}           & & \multicolumn{5}{c}{Tau1} \\
                                \cline{2-4}                              \cline{6-10} 
                                 & \multirow{2}{*}{AB}   
                                             & \multirow{2}{*}{AA}                                                         & \multirow{2}{*}{Traffic}
                                                                      & &  \multirow{2}{*}{}
                                                                                & \multicolumn{2}{c}{Mem. reduction}  & \multirow{2}{*}{Traffic}    & Traffic \\
                                                                                \cline{7-8}
                                 &           &           &            & &       &  AB            &  AA               &            &  reduction       \\
        \hline
        D2Q9                     & $18 s$    & $ 9 s$    & $18 s$     & & $6 s$ & $3    \times$  &  $1.5   \times$    &  $6 s$    & $3 \times$ \\
        D3Q15                    & $30 s$    & $15 s$    & $30 s$     & & $8 s$ & $3.75 \times$  &  $1.875 \times$    &  $8 s$    & $3.75 \times$ \\
        D3Q19                    & $38 s$    & $19 s$    & $38 s$     & & $8 s$ & $4.75 \times$  &  $2.375 \times$    &  $8 s$    & $4.75 \times$ \\
        D3Q27                    & $54 s$    & $27 s$    & $54 s$     & & $8 s$ & $6.75 \times$  &  $3.375 \times$    &  $8 s$    & $6.75 \times$ \\
    \end{tabular}
\end{table*}

An interesting feature of the Tau1 algorithm is the amount of data transferred per lattice node. In standard LBM, $2 \cdot q \cdot s$ bytes must be transferred per node. Tau1 requires reading the local velocities $\vec{u}$ and densities $\rho$ from the processed node and its neighbors ($q \cdot (d+1)$ values) and writing the local velocity and density for the currently processed node ($(d+1)$ values), resulting in $(q+1) \cdot (d+1) \cdot s$ bytes transferred in total. However, because the same local velocities and densities are accessed during the processing of neighboring nodes, we can assume that, for an ideal implementation, these values can be buffered in the processor's local memory (cache memory, registers, scratchpad memory). This assumption reduces the amount of data transferred from external memory to a single read and writes of $\vec{u}$ and $\rho$ per lattice node because other accesses to these data can use local copies stored inside the processor. Thus, for an ideal implementation of the Tau1 algorithm, we can define the size of the data transferred from external memory as $2 \cdot (d+1) \cdot s$ bytes per lattice node. 

A comparison of the memory usage and traffic of \mbox{Tau1} algorithm with standard LBM implementations is shown in Table \ref{tab_complexity}. Tau1 allows for significantly reducing both parameters, especially for lattice arrangements with multiple distribution functions. In particular, a decrease in memory traffic may lead to a proportional increase in performance, as LBM implementations are often bandwidth-bound. However, achieving low memory traffic requires storing copies of the velocities and densities for several neighboring nodes in the processor's internal memory (cache/scratchpad), which may increase the register pressure and require additional optimization techniques (e.g., cache line blocking). Moreover, because the computational complexity of the Tau1 algorithm is similar to that of the LBM, a significant decrease in the amount of transferred data increases the arithmetic intensity of the algorithm (the ratio of arithmetic operations to the size of the transferred data). Thus, the Tau1 algorithm may become compute-bound on a computer with a low ratio of computational power to available memory bandwidth (the machine processing capability). 

\section{Implementation} 
\label{implementation}

In this paper, we designed the optimized Tau1 implementations for the D2Q9, D3Q19, and D3Q27 schemes (the source code is available at \cite{Tomczak2023}). Our implementations are based on \cite{tau1} and written as templated code in C++, allowing for customizability while keeping the code relatively simple. All analyzed kernels were prepared as multithreaded codes with unrolled loops and SIMD processing and compiled with compiler flags enabling high level of code optimizations. SIMD processing was realized using Vector Extensions, available in many C++ compilers, to simplify code preparation and portability to different processor architectures (e.g., we could compile kernels for 3D schemes on a laptop with an Apple M2 processor). For 3D kernels, we enabled huge pages to decrease the impact of TLB (Translation Look-Aside Buffer) misses on performance. 

In addition to the general optimizations mentioned above, we also provided Tau1-specific optimization, significantly decreasing the number of division operations in the original Tau1 implementation \cite{tau1}. The algorithm described in \cite{tau1} stores the mass flux in the memory at each node and divides those fluxes by the density to reproduce the velocity at a given node. It is done for every direction, resulting in 81 divisions per node for D3Q27.

\begin{algorithm2e}[t!]
    \DontPrintSemicolon
    \ForAll{$\vec{x}$}
    {
        $\vec{u}(t, \vec{x}) \gets \vec{0}$\;
        $\rho(t, \vec{x}) \gets 0$\;
        \ForAll{k}{
            \eIf{$\vec{x}-\vec{c_k}$ is a fluid node}{
                $ik \gets i\;\;:\vec{c_{ik}}=-\vec{c_k}$\;
                $\vec{u} \gets \vec{u}(t-1, \vec{x}+\vec{c_{ik}})$\;
                $f^{eq}_{ik} \gets ...$\;
            }{...\tcp*[l]{fullway/halfway bounceback}}
            $\vec{u}(t, \vec{x}) \gets \vec{u}(t, \vec{x}) + e_{ik} \cdot f^{eq}_{ik}$\;
            $\rho(t, \vec{x}) \gets \rho(t, \vec{x}) + f^{eq}_{ik}$\;
        }
        $r_{\rho} = \frac{1}{\rho(t, \vec{x})}$\;
        $\vec{u}(t, \vec{x}) \gets \vec{u}(t, \vec{x}) \cdot r_{\rho}$\;
    }
    \caption{Optimized Tau1 algorithm that stores velocity at the node.}
    \label{tau1_div}
\end{algorithm2e}

In our work, we store the velocity instead of the mass flow rate in the node. Thus, the division is done for each node only once for each dimension. Additionally, since all coordinates of the velocity vector $\vec{u}$ are divided by the same $\rho$ value, then we can perform only a single computation of the reciprocal $r_{\rho} = 1 / \rho$ and then multiply $\vec{u}$ by $r_{\rho}$. For the D3Q27 lattice, this technique decreases the number of divisions 81-fold. A modified algorithm is presented in Algorithm \ref{tau1_div} with the reciprocal computations shown in lines 15 and 16.

\begin{figure}[!ht]
\centering
\includegraphics[width=0.99\columnwidth]{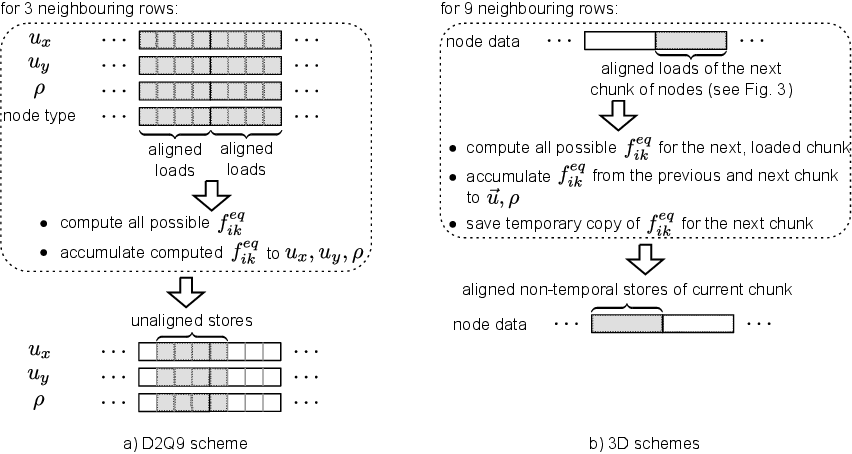}
\caption{
	The idea of vector processing of four neighboring nodes in a) D2Q9 and b) 3D schemes.
}
\label{fig_implementation}
\end{figure}

The optimized kernels were prepared in two stages. In the first stage, we implemented a simple vectorized version of the D2Q9 scheme. This kernel achieved about 70 \% of theoretical memory bandwidth on one of our systems but resulted in low performance on the other system (see details in Section \ref{sec_performance}). In the second stage of code preparation, we implemented kernels for 3D schemes with additional optimizations: non-temporal memory stores, more local data memory layout, and better overlapping of memory operations with computations.

\subsection{Kernel for D2Q9}

The general idea of neighboring nodes processing in the D2Q9 kernel is shown in Fig. \ref{fig_implementation}a.
We process $V_s = 4$ neighboring nodes in parallel, where $V_s$ denotes the width of vector data types.
Processing four nodes simultaneously allows for using x86 AVX (Advanced Vector Extensions) instructions and double-precision data format. The processed nodes are from the same row; for example, from $(x, y)$ to $(x+V_s-1, y)$ where $x, y$ denote the column and row coordinates, respectively, and $(x-1)$ is divisible by $V_s$. The memory layout uses two copies of data (AB layout) and row-major node ordering with separate arrays per components of $\vec{u}$ and $\rho$ (Structure-of-Arrays), allowing loading of the velocity and density of $V_s$ neighboring nodes to a vector register with a single instruction. Additionally, we guarantee that all memory reads will be aligned, enabling the usage of faster memory transfer instructions, such as {\tt vmovapd} for the AVX instructions.

During the processing of $V_s$ neighboring nodes, we need access to values from neighboring nodes at $x-1$ and $x+V_s$ columns, as well as from rows placed "above" and "below" the currently processed nodes (rows at $y+1$ and $y-1$). To decrease the number of memory transfers, we first load to vector registers the values from $2 \cdot V_s$ neighboring nodes from a single row, starting from a node at column $x-1$. Then, we use the {\tt \_\_builtin\_shufflevector()} function to extract the proper values during the computations of functions $f_{ik}^{eq}$. 
We also prepared the manually unrolled loop over the neighboring directions with changed order of neighbor node processing, allowing to process of all neighbor nodes from a single row (e.g., nodes placed at directions $e_i \in \left\{(-1,e_y,e_z), (0,e_y,e_z), (1,e_y,e_z,)\right\}$) after a single load of values for $2 \cdot V_s$ neighboring nodes.%

The implementation also contains separate paths in the source code for cases where not all $V_s$ neighboring nodes are fluid (for example, for the boundary nodes close to the geometry walls). 
When the $V_s$ neighboring nodes contain at least one non-fluid node, we compute two versions of $f^{eq}_{ik}$ functions (fluid and boundary) for all neighbor nodes (because we use vector processing) and combine the resulting vector using boolean operations.
The above procedure allows us to avoid scalar code (at the cost of additional computations) and minimize costly branches (using boolean masks).

\subsection{Kernel for 3D schemes}
\label{sec_kernel_3d}

The performance analysis of the D2Q9 kernel implementation (presented in Section \ref{sec_performance}) showed that the main reason for limiting performance was additional memory traffic caused by the memory write-allocation policy.
Additionally, we observed additional performance penalties for the system based on AMD Ryzen 9 7945HX CPU when neighboring memory operations (especially stores) referred to distant addresses.  
In kernels for 3D schemes, we then mitigated these issues.

\begin{figure}[!ht]
\centering
\includegraphics[width=0.99\columnwidth]{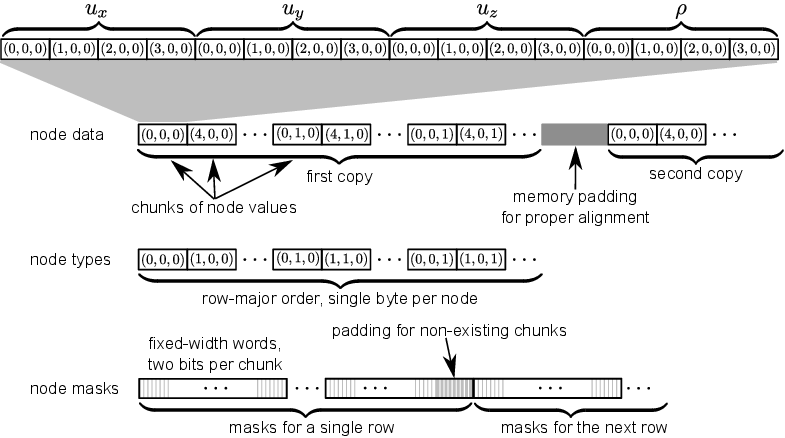}
\caption{Memory layout used in vector implementations for 3D schemes. The coordinates of nodes are in $(x,y,z)$ order.}
\label{fig_mem_layout}
\end{figure}

The main improvement in 3D kernels (see Fig. \ref{fig_implementation}b) is that computed values of $\vec{u}$ and $\rho$ for $V_s$ neighboring nodes can be stored with aligned vector operation allowing for usage of non-temporal instructions. The kernels for 3D schemes were prepared with a different memory layout (shown in Fig. \ref{fig_mem_layout}) where a single copy of all values for neighboring $V_s$ nodes is stored in neighboring memory locations (called "chunk"), which is properly aligned for vector loads/stores. To decrease the number of memory operations on \texttt{node\_data} array containing values of velocity and density for nodes, in a single iteration of the loop over neighboring chunks, we load only one next chunk from processed rows and store the loaded values in local variables (see Fig. \ref{fig_implementation}b). This method gives a few percent increase in performance due to more instruction-level parallelism and partial hiding of memory latencies with computations on data loaded during the processing of the previous chunk. One of the downsides of this approach is that it leads to higher register pressure, which in turn causes an increase in the number of register spills that occur, especially for schemes with multiple neighbor directions. Fortunately, the data spilled to the stack can usually be stored in cache memory, decreasing its negative impact on performance. 

For 3D kernels, we also used an additional \texttt{node\_mask} array that uses two bits per chunk to assign a chunk one of the three types: a chunk with solid nodes only (solid chunk), a chunk with fluid nodes only surrounded by fluid nodes only (fluid chunk), and chunk with a mix of fluid and solid nodes and/or with non-fluid neighbor nodes (mixed chunk). This information allows for skipping loading node types for fluid chunks and completely skipping the processing of solid chunks. Thus, we provided two paths in chunk processing code: the highly-optimized version (shown in Fig. \ref{fig_implementation}b) for fluid chunks and a separate version for mixed chunks where node types must be checked to process boundary nodes as in the D2Q9 version. 

In code for mixed chunks, we resigned from overlapping computations with data loads for the next chunk because mixed chunks are often adjacent to solid chunks (in rare situations only, mixed chunks are placed between other mixed or fluid chunks and are not adjacent to solid chunks). This decision allowed for simplification of the code at the cost of some performance penalty caused by more visible latencies of memory operations and additional instructions required for proper data preparation for the fluid chunk following the mixed chunk.

\subsection{Validation}

In order to validate our code, we first verified our optimized Tau1 implementation in two dimensions in two standard benchmark cases: the Poiseuille and lid-driven cavity flows. First, we ran the Poiseuille flow in the channel of width $H$ driven by a constant pressure gradient set up on a $1000\times 100$ grid with no-slip walls at the top and bottom. We compared our numerical results with the analytical solution \cite{LBM_principles}: 
            \begin{equation}
            \label{poiseuile}
                u_x(y) = -\frac{1}{2\rho\nu}\frac{\partial p}{\partial x}y(y-H),
            \end{equation}
where $\nu$ is fluid viscosity and $p$ is the pressure. 
We used the halfway bounce-back condition for no-slip and constant pressure boundary at the inlet with $\rho_{in} = 1.01$ and outlet with $\rho_{out} = 0.99$. The simulations were run until they reached a steady state, defined by the condition 
$ \left|\vec{u}(\vec{x}, t) - \vec{u}(\vec{x}, t-1)\right| < \varepsilon,$
where $\varepsilon=10^{-8}$ is evaluated in subsequent time steps in each node of the entire grid.
With this condition, the simulation converged after approximately 200000 iterations (convergence was checked every 10000 iterations). Then, the fluid velocity was captured at $x=500$.
The mean absolute percentage error between our simulation and analytical solution was $\mathrm{MAPE}\approx 0.084\%$.
            
\begin{figure}[!ht]
\centering
\includegraphics[width=0.6\linewidth]{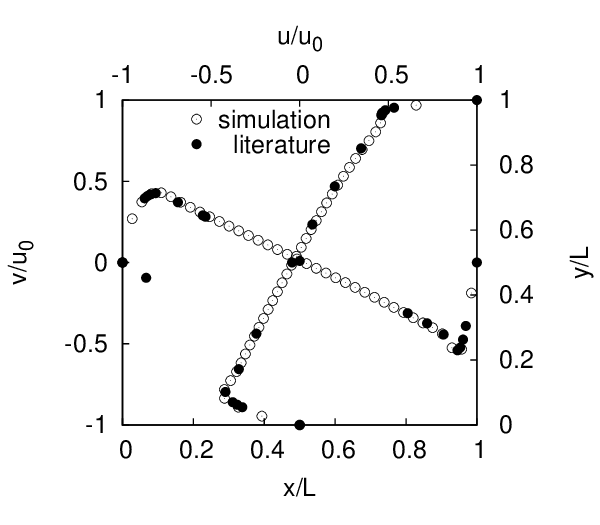}
\caption{The centerline profiles of velocity in the lid-driven cavity at Re=3200 computed with optimized Tau1 implementation compared to benchmark data from \cite{Ghia1982}.}
\label{re3200_cl}
\label{poiseuille_profile}
\end{figure}

Then, we run the lid-driven cavity benchmark with the top boundary moving with a constant velocity $u_x = \frac{5}{12}$ $lu/ts$. The remaining walls were set as halfway bounce-back no-slip. The Reynolds number was $Re=2.5\times L$, where L was the grid size. We performed finite-size scaling tests for $L=160, 400$ and $1280$ resulting in $\mathrm{Re}=400,1000$ and $3200$, respectively. Here, the relative changes in the volumetric flow rate through a vertical cross-section were initially checked for convergence every 1000 time steps. The simulation was stopped at a steady state when changes were smaller than $1\%$ for 10000 time steps (checked every 100 time steps). 
We gathered the horizontal velocity $u$ at the vertical cross-section ($x=L/2$) and 
the vertical velocity $v$
at the horizontal cross-section ($y=L/2$) (see Fig. ~\ref{re3200_cl}). 
We found that profiles are in perfect agreement with the literature.
\subsection{The influence of optimization on the algorithm in 3D}

\begin{figure}[!ht]
\centering
\includegraphics[width=0.99\linewidth]{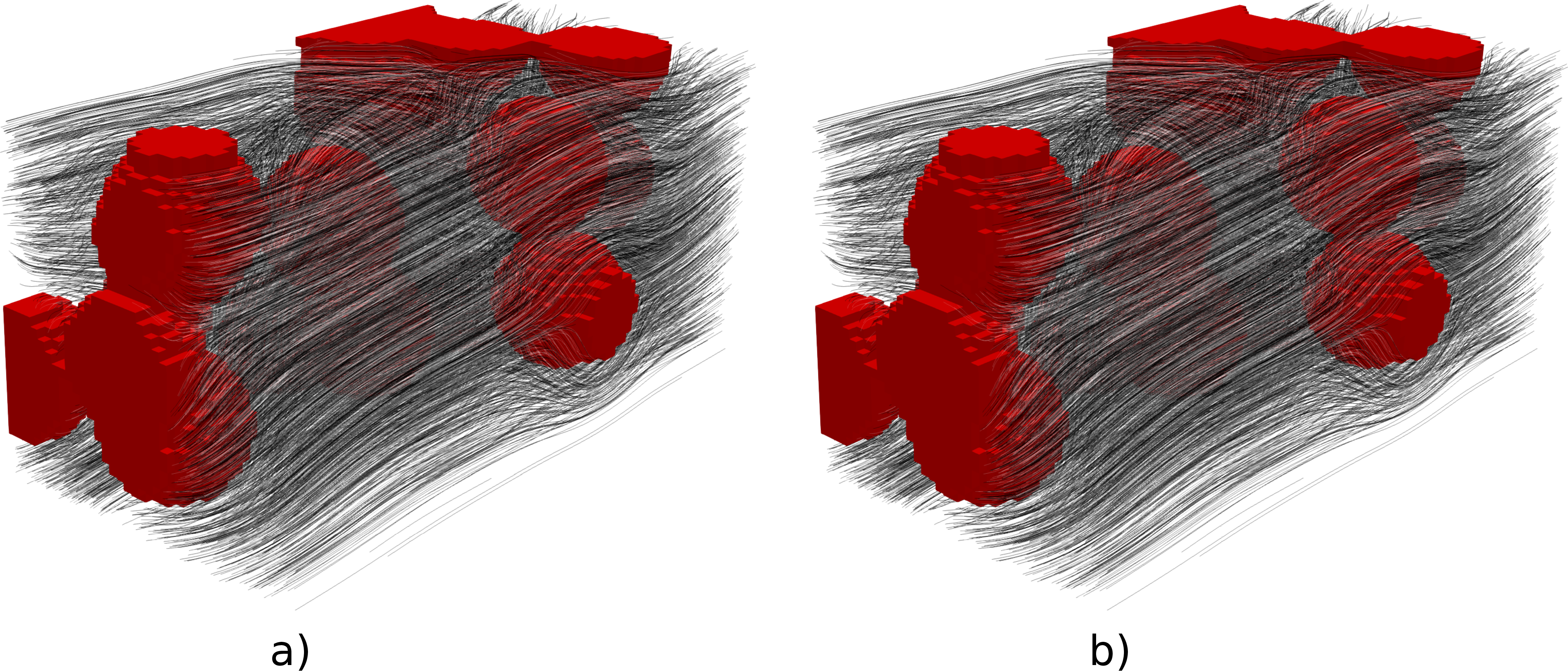}
\caption{Direct comparison between fluid flow results for a 3D system of flow penetrating pore scale porous medium computed using a) reference and b) highly optimized Tau1 code. The solid (red in color) obstacles in the flow are blocking the fluid. Streamlines are plotted as black and white lines.}
\label{fig3d1}
\end{figure}

Next, we checked and verified the influence of optimizations on the results of three-dimensional flows. We directly compared standard reference results (unoptimized code) with highly optimized Tau1 implementation in the fluid flows through randomly packed porous samples at varying porosity in three dimensions. As a first comparison, one can visually compare fluid flow streamlines in this system, which, in our case, do not differ between the results of optimized and unoptimized kernels (see Fig.~\ref{fig3d1} for reference).
To numerically compare velocity fields in relative differences, we compared velocity fields between both implementations by taking the difference between results taken at the 10000 time step. The difference was calculated node by node, and the maximum absolute difference and the maximum relative differences among all grid points for each velocity component were found. 
We found, for double precision, that differences between the reference and optimized kernels appear to be negligibly small, with at least seven significant decimal digits agreement on the maximum velocity (max absolute difference is $10^{-12}$, and max velocity is $10^{-5}$).
For single precision, the differences are at the second significant decimal digit for D3Q19 and D3Q27.

\section{Performance analysis}
\label{sec_performance}

Performance was measured on two systems: one based on Intel Core i9-12900K processor (8 performance cores set at 4.9 GHz and eight efficient cores at 3.7 GHz), and one based on AMD Ryzen 9 7945HX (16 cores at 5 GHz), respectively.
On both processors, hyperthreading was enabled.
For i9-12900K, we used the gcc-12.2.0 compiler on the Linux Ubuntu 22.04 system with a 5.15.0-52-generic kernel.
For 7945HX, the AMD clang version 14.0.6 compiler and Linux Ubuntu 23.04 with 6.2.0-34-generic kernel were installed.

Both systems use dual-channel DDR5 4800 MT/s memories with a theoretical peak bandwidth of 76.8 GB/s. However,  benchmarks (Intel VTune for i9-12900K and Aida64 for 7945HX) showed significant differences in sustainable memory bandwidth: the system based on i9-12900K achieved 68 GB/s whereas, for 7945HX, only 45 GB/s was reported during memory copy. The possible causes of the low performance of the system based on 7945HX are single-rank, mediocre latency memory modules with a low number of banks ($2 \times 8$ GB SODIMM Micron MTC4C10163S1SC48BA10) and microarchitectural limits of 7945HX processor (e.g., delays in the connection between processor cores and memory controller). Unfortunately, the detailed analysis is complicated due to incomplete information about performance counters available in Data Fabric and Unified Memory Controller \cite{AMD_PPR}. In our case, especially important may be limitations for memory writes as presented in \cite{AMD_SOG}: decreased throughput when the number of write-combining streams grows, conflicts during store-to-load forwarding for memory operations with the same 12 least significant address bits, and halved throughput for memory write instructions.

The performance of LBM algorithms is given in fluid lattice updates per second (FLUPS) equal to the number of non-solid nodes processed during one second of the wall clock time measured using a high-resolution clock from the \texttt{std::chrono} library. Additionally, we provide the theoretical memory bandwidth based on the data from Table \ref{tab_complexity}. The performance measurements were done for 1000 time step iterations of the given simulation. Before measurements, we first launched 2000 iterations for each simulation to "warm up" the processor (fill up cache memories and stabilize die temperature) before the measurements.

\subsection{Performance for D2Q9 scheme}

\begin{figure}[!ht]
\centering
\includegraphics[width=0.45\columnwidth]{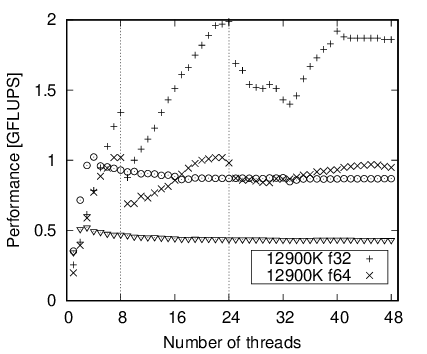}
\includegraphics[width=0.45\columnwidth]{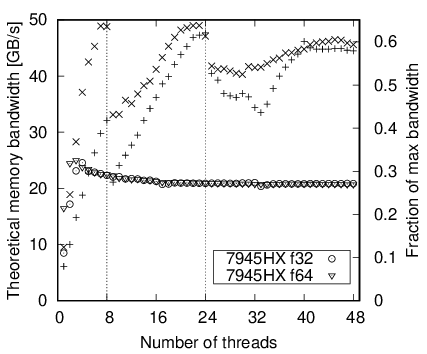}
\includegraphics[width=0.45\columnwidth]{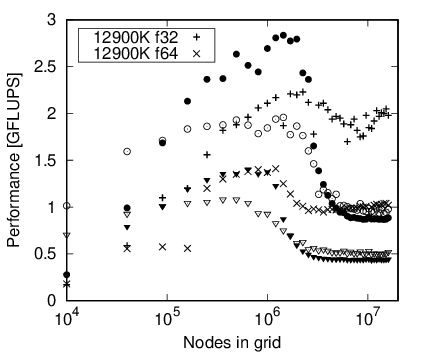}
\includegraphics[width=0.45\columnwidth]{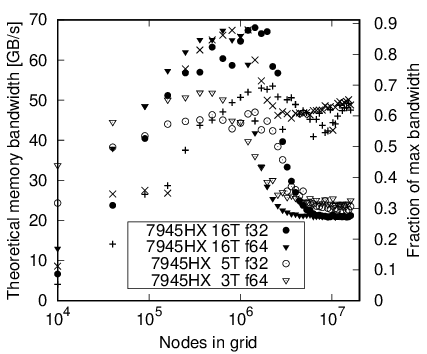}
\caption{
		Performance of Tau1 optimized D2Q9 kernels.
		In the top row of plots, the results are for dense geometry with $4000^2$ nodes. 
		In the bottom row of plots, 23 threads were used for i9-12900K, whereas the number of threads for 7945HX is given in legend.
		Memory traffic is computed from Table \ref{tab_complexity}.
		The fraction of memory bandwidth is computed for a maximum of 76.8 GB/s. 
		Symbols are the same on both plots from the same row.
		"f32" and "f64" denote single- and double-precision.
 }
\label{fig:perf2D}
\end{figure}

First, the performance measurements for D2Q9 kernels were done on both systems with the results presented in Fig. \ref{fig:perf2D}. 

For geometries occupying more memory than cache capacity, the Tau1 performance corresponds to about $2/3$ of the theoretical peak memory bandwidth for the system based on i9-12900K and to about $1/3$ for 7945HX (for a small number of threads). Since the actual amount of data transferred in our D2Q9 kernel is about $1.5\times$ larger than data presented in Table \ref{tab_complexity} due to write-allocation memory policy, the actual memory bandwidth is equal to about 100\% of memory bandwidth reported by VTune for i9-12900K and to about 80\% of Aida64 benchmark on 7945HX.

We observe significant differences in the performance behavior of the two systems (Fig.~\ref{fig:perf2D}). 7945HX processor allows for much higher Tau1 performance per single core but only when the memory bandwidth is not fully utilized, i.e., for a small number of threads or small cases fitting cache memory. Moreover, in the bandwidth-limited area, the performance of 7945HX decreases with a growing number of threads. In contrast, the performance of i9-12900K steadily grows with an increasing number of threads (despite a sudden drop in performance at nine threads when low-performance efficient cores are used). Since the sustainable memory bandwidth of the system based on 7945HX is much lower than for the system based on i9-12900K, the latter system achieves up to two times higher performance for geometries containing at least millions of nodes. The system based on 7945HX has a significant advantage only for small geometries and single-precision computations. Notice also that the performance for 7945HX starts decreasing for  smaller cases than for i9-12900K despite larger cache capacity (e.g., $2 \times 32$ MB L3 for 7945HX compared to 30 MB for i9-12900K), most probably due to additional latencies in 7945HX during access to L3 cache from the other Core Complex Die (CCD).

\subsection{Performance in 3D}

\begin{figure}[!ht]
\centering
\includegraphics[width=0.45\columnwidth]{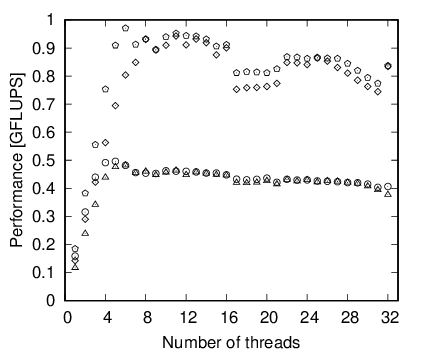}
\includegraphics[width=0.45\columnwidth]{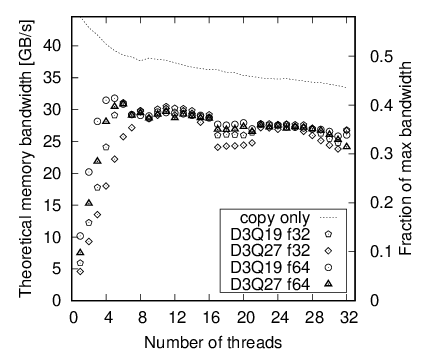}
	\caption{
		Performance of Tau1 optimized kernels on 7945HX processor as a function of the number of threads for dense geometry with $512 \times 256 \times 256$ nodes. 
		Memory traffic is computed from Table \ref{tab_complexity}. 
		The fraction of memory bandwidth is computed for the theoretical maximum memory bandwidth of dual-channel 4800 MT/s DDR5 memory (76.8 GB/s). 
		Symbols are the same on both plots.
		"f32" and "f64" denote single- and double-precision floating-point data.
 }
\label{fig:perf3D_nThr}
\end{figure}

The performance of Tau1 kernels for three-dimensional lattice arrangements was measured for the 7945HX-based system only (we lost access to i9-12900K), and the results are shown in Fig. \ref{fig:perf3D_nThr}. 
For large geometries, similarly to the D2Q9 kernel, the obtained performance is bounded only by the performance of the memory subsystem. Regardless of lattice arrangements and data types (f32/f64), all kernels achieve over 30 GB/s (39 \% of the theoretical maximum bandwidth and 67 \% of the measured maximum bandwidth equal to 45 GB/s). However, the maximum performance is obtained for different number of threads - the higher number of threads is needed for kernels with the higher arithmetic intensity (5 threads - D3Q19 f64, D3Q27 f64; 6 threads - D3Q19 f32; 11 threads - D3Q27 f32). After these peaks, the performance decreases with an increasing number of threads, probably due to multiple interleaved data streams stored in memory. We cannot completely confirm this hypothesis without a detailed analysis of microarchitecture bottlenecks. However, it seems consistent with \cite{AMD_SOG} and with our microbenchmarks that showed significant bandwidth decrease, even for a single thread, with increasing interleaved data streams stored in the memory. Additionally, the bandwidth for a simple data copy kernel (the loop with two 64-byte aligned memory reads and two 64-byte non-temporal memory writes per iteration) also drops with the number of threads.

For single thread, the lattice arrangement has significant impact on the performance which is equal to 159 MFLUPS (10.2 GB/s) and 111 MFLUPS (7.1 GB/s) for double-precision D3Q19 and D3Q27 kernels, respectively, and to 185 MFLUPS (5.9 GB/s) and 143 MFLUPS (4.6 GB/s) for single-precision D3Q19 and D3Q27 kernels. Notice that single-precision kernels achieve lower bandwidth than double-precision versions; thus, the single-thread performance is bounded by the number of instructions inside the kernels. 
The number of retired instructions per single fluid chunk of nodes is about 150 for D3Q19 and about 220 for D3Q27, including branches, general purpose and vector instructions, and intensive register spilling (about 40\% of kernel instructions require memory access, usually on the stack).

During measurements, we observed an issue caused by memory traffic due to register spilling. Sporadically, especially for double-precision D3Q27, and when the number of threads was divisible by 3, the performance of our kernels dropped several times, and we observed an increase in L2 TLB and cache misses. We have not found a simple, reliable way to avoid such behavior (the best would be manual management of register usage and spilling to properly aligned memory locations) and decided to restart the application in these cases because such performance drop was not frequent and can be easily detected at the beginning of the simulation.

\subsection{Performance for porous geometries}

\begin{figure}[!ht]
\centering
\includegraphics[width=0.45\columnwidth]{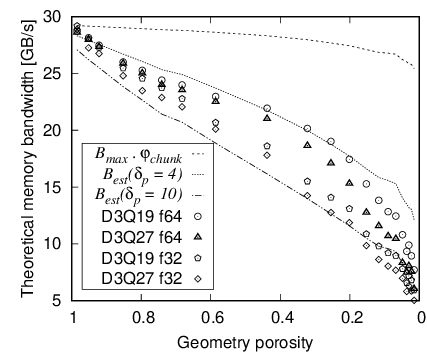}
\includegraphics[width=0.45\columnwidth]{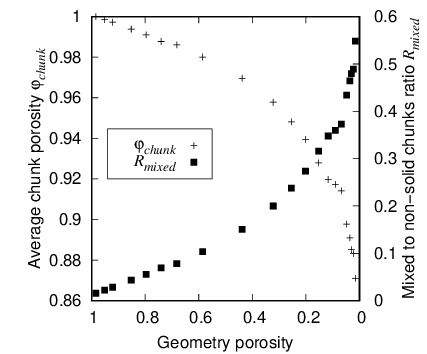}
	\caption{
		Performance of Tau1 optimized kernels on 7945HX processor (plot on the left) and geometry parameters (plot on the right) for geometries with $512 \times 256 \times 256$ nodes and different numbers (0 to 300) of randomly placed spherical obstacles.
  		Memory traffic is computed from Table \ref{tab_complexity}, considering only fluid nodes.
		"f32" and "f64" denote single- and double-precision floating-point data.
		All measurements were done for 16 threads.
  The percolation threshold $p_c$ of the system of overlapping spheres is $p_c\approx 0.0317$ according to \cite{Priour2014}.
 }
\label{fig:perf3D_porous}
\end{figure}

The presented kernels and data structures are optimized for dense geometries, but we also use \texttt{node\_mask} array, allowing us to skip computations for solid chunks (see Section \ref{sec_kernel_3d} for details). The performance evaluation for geometries with a wide range of porosities $\phi \geq 0.015$, equal to the ratio of the number of fluid nodes to all nodes, is shown in Fig. \ref{fig:perf3D_porous}. 

The main reason for the performance drop with decreasing porosity is an irregular memory access pattern for geometries with many obstacles. Detailed profiling showed that, for geometries with the lowest porosities, the number of accesses to the external DDR memory and the last level cache (including cache from the other CCD) is 1-2 orders of magnitude higher than for geometries without obstacles. Some penalties are also caused by less efficient code for mixed chunks (the code contains about three times more instructions and some redundancy) but are not dominant. When we prepared geometry with all chunks set to mixed type (by setting a single solid node per chunk), the chunks were processed at almost full speed (we observed 20-21 GB/s for $\phi = 0.75$, which gives 27-28 GB/s including data transferred for solid nodes from processed chunks).

A simplified performance model can also show the impact of latency penalties caused by irregular data arrangement for porous geometries. Assume that, for some porous geometry, there is always a large number of solid nodes before an initial chunk, i.e., the mixed chunk placed at the beginning of several fluid nodes. Notice that only about half of the mixed chunks are initial chunks; the other mixed chunks are at the end of a series of fluid chunks or between two neighboring fluid (or mixed) chunks. The average time needed to transfer the chunk data for such geometry can be then estimated as $t_{avg} = t_{min} + t_p \cdot R_{mixed} /2$, where $t_{min}$ denotes the time for transferring chunk data with maximum bandwidth $B_{max}$, $t_p$ is some additional latency penalty (caused by irregular memory access pattern for initial chunks), and $R_{mixed}$ denotes the ratio of the number of mixed chunks to the number of all transferred, non-solid chunks (mixed and fluid). Assuming that $M_{chunk}$ denotes the amount of data transferred for a chunk (mixed or fluid), we can then define the ratio of the maximum bandwidth $B_{max} = M_{chunk}/t_{min}$ to the average bandwidth for chunk data transfers $B_{chunk} = M_{chunk} / t_{avg}$ as $B_{max} / B_{chunk} = t_{avg} / t_{min}$ what allows to find $B_{chunk} = B_{max} / (1 +  \delta_p \cdot R_{mixed} /2)$  where $\delta_p = t_p / t_{min}$. Eventually, since not all nodes in chunks are fluid, the bandwidth estimated for fluid nodes is 
\begin{equation}
	B_{est} = B_{chunk} \cdot \phi_{chunk} = \frac{\phi_{chunk} \cdot B_{max}}{ 1 +  \delta_p \cdot \frac{R_{mixed}}{2}}
	,
\end{equation}
where $\phi_{chunk}$ denotes the average porosity of non-solid chunks.

The example values of $B_{est}$ for different $\delta_p$ are shown in Fig. \ref{fig:perf3D_porous}; the values of $\delta_p$ are close to the actual ratio for DDR5 memories, e.g. single burst transfer requires 16 memory bus cycles whereas the latency penalty caused by a sequence of row precharge, activation and column access takes about 40+39+39 cycles for MTC4C10163S1SC48BA10 memory. The general shape of $B_{est}$ resembles the measured bandwidth, but $B_{est}$ is not exact because of oversimplifications. For example, in complex geometries, not all initial chunks are preceded by many solid nodes, forcing memory row precharge. 

Fig. \ref{fig:perf3D_porous} also contains bandwidth estimation resulting from chunk porosity only $B_{max} \cdot \phi_{chunk}$.
The performance drop from processing solid nodes in mixed chunks is small because the average chunk porosity stays high (over 0.85) even for $\phi= 0.015$ due to the small chunk size (4 nodes per chunk).
Notice that mixed chunks constitute less than half of all non-solid chunks except for the most porous geometries.

\subsection{Performance comparison with other LBM implementations}

We compare the proposed implementation of the Tau1 algorithm with other high-performance implementations of classical LBM (see Tab.~\ref{tab:perf_comp}). We selected implementations with the highest performance from the referenced papers. For each implementation, we provide the raw performance measured in MFLUPS and additional parameters related to the memory bandwidth of used machines (summarized in Table \ref{tab:cpus}) that simplify the comparison of different algorithms on different hardware.

\begin{table*}[t]
    \centering
    \caption{        
			Memory bandwidth (in GB/s) for processors used in LBM performance evaluation. For systems marked with $^\dagger$, the actual bandwidth was not given and assumed based on similar hardware.
    }
    \label{tab:cpus}
    \begin{longtable}{c c c c }
         CPU                         & Max bandwidth  & Real bandwidth  & Real/max ratio \\
        \hline                                                                                                                                          
        Xeon E5-2695v3                       &  68.3  &  52     & 0.762 \\
        Xeon Gold 6148                       & 128    &  102.8  & 0.803 \\
        EPYC 7451                            & 170.6  &  130.9  & 0.767 \\
        Core i9-10980XE                      & 93.9   &  65     & 0.693 \\
        $4\times$ Xeon E5-4620 v4            & 273    &  126    & 0.461 \\
				$2\times$ Xeon E5-2697 v3$^\dagger$  & 136.5  &  104    & 0.762 \\
				NEC SX-ACE$^\dagger$                 & 256    &  $\approx$220  &  $\approx0.86$     \\
				$5\times$ Xeon E5-2690 v3$^\dagger$  & 341.3  &  $\approx$256  &  $\approx0.75$     \\
				SW26010                              & 136.5  & 128     & 0.938 \\
        Core i9-12900K                       & 76.8   &  68     & 0.807 \\
        Ryzen 9 7945HX                       & 76.8   &  45     & 0.586 \\
    \end{longtable}
\end{table*}

\begin{table*}[tbh]
    \centering
    \caption{
        Performance in MFLUPS (given in "Perf." column) of different implementations. 
				The performance of this work is given for large geometries ($4000^2$ for D2Q9 and $512\times256\times256$ for 3D schemes) and the number of threads resulting in the best performance.
        The $U_B$ column shows the LBM implementation's memory bandwidth ratio to the processor's measured memory bandwidth.
				The last two columns contain the ratios of the lattice node updates (MFLU/GB) and particle distribution function updates (MPDU/GB) to the measured memory bandwidth.
    }
    \label{tab:perf_comp}
    \begin{tabular}{c c c c c c c}
                                & CPU                & LBM        & Perf.   & $U_B$  &  MFLU/GB    & MPDU/GB \\
        \hline                                                                                                                                          
        \cite{Bauer:lbm2021}    & E5-2695v3          & D3Q19 f64  &  ~150   & 0.877  &   ~2.9     &  ~55 \\
                                & Xeon 6148          & D3Q19 f64  &  ~304   & 0.899  &   ~3.0     &  ~56 \\
				\cite{Latt2021Pal}      & $5\times$ E5-2690  & D3Q19 f64  &  ~135   & 0.160  &   ~0.5     &  ~10 \\
				                        & 7945HX             & D3Q19 f64  &  $\ \ 58$   & 0.392  &   ~1.3     &  ~24 \\
        \cite{Lehman:Acc2022}   & i9-10980XE         & D3Q19 f64  &  ~110   & 0.515  &   ~1.7     &  ~32 \\
                                & $4\times$ E5-4620  & D3Q19 f64  &  ~151   & 0.364  &   ~1.2     &  ~23 \\
        \cite{Liu2019Sun}       & SW26010            & D3Q19 f64  &  ~281   & 0.668  &   ~2.2     &  ~42 \\
				\cite{Qi2016Perf}       & $2\times$ E5-2697  & D3Q19 f64  &  ~155   & 0.453  &   ~1.5     &  ~28 \\
				                        & SX-ACE             & D3Q19 f64  &  ~310   & 0.428  &   ~1.4     &  ~27 \\
        \cite{Wittmann:Lat2018} & EPYC 7451          & D3Q19 f64  &  ~325   & 0.755  &   ~2.5     &  ~47 \\
        \textbf{this}           & i9-12900K          & D2Q9  f64  &  1020   & 0.720  &   15.0     &  135 \\
        \textbf{work}           &                            & D2Q9  f32  &  1981   & 0.699  &   29.1     &  262 \\
                                & 7945HX             & D2Q9  f64  &  ~521   & 0.556  &   11.6     &  ~96 \\
                                &                    & \textbf{D3Q19 f64}  &  \textbf{~497}   & \textbf{0.706}  &   \textbf{11.0}     &  \textbf{210} \\
                                &                    & D3Q27 f64  &  ~482   & 0.686  &   10.7     &  289 \\
                                &                    & D2Q9  f32  &  1023   & 0.546  &   22.7     &  113 \\
                                &                    & D3Q19 f32  &  ~971   & 0.690  &   21.6     &  410 \\
                                &                    & D3Q27 f32  &  ~943   & 0.671  &   21.0     &  566 \\
    \end{tabular}
\end{table*}

Bandwidth utilization $U_B$ comes directly from the Roofline performance model \cite{Williams:Roof2009} and equals to the ratio of the theoretical data traffic (calculated as the performance in MFLUPS multiplied by the data transferred per node as presented in Section \ref{sec:complexity}) to the measured memory bandwidth. Next, because our implementation requires much lower memory traffic per node than the reference implementations, we use the second parameter computed as the ratio of performance in MFLUPS to the maximum measured memory bandwidth and measured in million lattice updates per gigabyte (MFLU/GB). To better compare implementations with different lattice arrangements, we also use the performance measured in millions of particle distribution function updates per gigabyte (MPDU/GB) and computed as a ratio of performance measured in millions of particle distribution function updates per second to the memory bandwidth.

The performance measured in processed distribution functions, instead of lattice nodes, can be a reasonable method of comparison of implementation efficiency for different lattice arrangements. According to \cite{Bauer:lbm2021}, the arithmetic complexities of compressible BGK LBM are 90, 193, and 293 FLOPs for the D2Q9, D3Q19, and D3Q27 lattice arrangements, respectively, giving approximately 10 FLOPs per distribution function regardless of the lattice arrangement. Thus, for the presented cases, the computational complexity and the amount of transferred data of LBM kernels scale linearly with the number of velocity directions. The direct comparison of the processing speed of a single distribution function allows for comparing different LBM implementations for different lattice arrangements. For example, processing a single distribution function requires transferring 16 bytes for double-precision data. The theoretical maximum speed of processing distribution functions equals 62.5 MPDUS per one GB/s of available memory bandwidth. This parameter is the constant limit for standard LBM implementations. It defines the maximum MFLUPS and MFLU/GB performance for bandwidth-bound implementations on machines with a given memory bandwidth. Similar reasoning (but with a performance limit of about 100 MPDUS per 1 GFLOPS) applies to machines where the implementation is computation-bound.

The data from Table \ref{tab:perf_comp} shows that, for the D3Q19 double-precision version, Tau1 implementation achieves at least $497 / 325 = 1.5\times$ higher performance (measured in MFLUPS) than highly optimized, state-of-the-art implementations of standard LBM, even though we use a CPU with $2.9\times$ lower memory bandwidth than EPYC 7451. Thus, in comparison to the actual memory bandwidth of the system, Tau1 achieved $4.4\times$ higher performance measured in MFLUPS.
A more detailed comparison of performance concerning real memory bandwidth can be done based on the last columns from Tab. \ref{tab:perf_comp}. Double-precision Tau1 implementation for 3D schemes allows to process close to 11 MFLUPS per one GB/s of available memory bandwidth whereas the standard LBM implementations achieve up to 3 MFLU/GB. Notice that the maximum theoretical performance for double-precision standard LBM is 3.29 MFLUP/GB because computations for a single node require transferring 304 bytes. For double-precision Tau1 and 3D schemes, the theoretical upper limit is $1~[\mathit{GB}] / 64~[\mathit{bytes~per~node}] = 15.6$ MFLUP/GB. The additional advantage of Tau1 is its almost constant MFLUPS performance regardless of the number of velocity directions, resulting in very high performance in MPDU/GB. Standard LBM implementations are bounded by the processing speed of a single distribution function (up to 56 MPDU/GB in Table \ref{tab:perf_comp}) and, as a result, their MFLUPS performance drops for lattices with a larger number of links.

\section{Conclusion and future work}

We have presented and evaluated the high-performance implementation of Tau1, a recently developed modification of the standard lattice-Boltzmann algorithm. Our analysis has shown that the Tau1, although designed to decrease memory usage, significantly changes the algorithm performance characteristics and enables a several-fold increase in performance compared to the standard LBM. For double-precision D3Q19 implementation, Tau1 resulted in at least $3.67\times$ higher performance compared to sustained memory bandwidth than highly-optimized standard LBM implementations: Tau1 achieved $11 \cdot 10^6$ fluid node updates per 1 GB/s (MFLU/GB) of available memory bandwidth (497 MFLUPS on 7945HX system with 45 GB/s during memory copy) whereas standard LBM allows up to 3 MFLU/GB (304 MFLUPS on Xeon Gold 6148 system with 102.8 GB/s actual memory bandwidth). The advantage of Tau1 is also almost constant performance measured in fluid lattice updates per second regardless of the number of velocity directions. In contrast, the performance of standard LBM decreases with an increasing number of lattice links.

The high-performance Tau1 implementation poses different challenges than standard LBM. The computational complexity of Tau1 is comparable to standard LBM, but Tau1 requires buffering of velocities and densities from neighboring nodes due to the intensive reuse of these data. Moreover, in theory, Tau1 BGK may result in low register pressure due to the single use of the equilibrium functions for the neighboring nodes. However, in our system with slow, high latency memory and bandwidth penalties for multithreaded memory writes, we had to prepare the code with a lot of data buffering, allowing for increasing instruction level parallelism and partial hiding of memory latencies at the cost of intensive register spilling. Another characteristic of Tau1 that distinguishes it from standard LBM is a higher number of arithmetic operations per single byte of transferred memory (the arithmetic intensity) resulting from decreased memory traffic. It may cause Tau1 implementations to become computation-bound on machines with low machine balance. However, on the system based on a 7945HX processor with 43 double-precision FLOP per byte of actual memory bandwidth, we saturated memory with 5 to 11 threads (5 for double-precision D3Q19 and D3Q27 kernels, 11 for single-precision D3Q27).

For 3D schemes, the performance evaluation was done on AMD Zen 4 architecture; thus, we used the memory layout that stores velocity and density for chunks of neighboring nodes in consecutive memory locations to decrease the number of simultaneous memory write streams. Although this layout is optimized for dense geometries, we also applied limited support for sparse geometries by using an additional bitmask enabling different paths in code for node chunks containing only solid, only fluid, and a mix of solid and fluid nodes. The performance evaluation for sparse geometries showed that, for low porosities, the main reason for low performance is an irregular memory access pattern, making it challenging to use cache memory effectively and exposing memory latencies.

Further work includes research for techniques of efficient implementation for sparse geometries. 
It may require a different approach than standard LBM due to both different memory access patterns and increased memory overheads caused by additional data for handling geometry sparsity (since Tau1 requires less memory per node data, additional data for sparsity handling become more visible).
We are also planning to search for optimizations to reduce register pressure and the number of instructions per kernel and increase memory bandwidth usage.

\section*{Acknowledgments}
Funded by National Science Centre, Poland under the OPUS call in the Weave programme 2021/43/I/ST3/00228. This research was funded in whole or in part by National Science Centre (2021/43/I/ST3/00228). For the purpose of Open Access, the author has applied a CC-BY public copyright license to any Author Accepted Manuscript (AAM) version arising from this submission. Created using resources provided by Wroclaw Centre for Networking and Supercomputing (http://wcss.pl).

\bibliographystyle{unsrt}


\begin{thebibliography}{10}

\bibitem{Gmeiner2015}
Bjorn Gmeiner, Ulrich R{\"u}de, Holger Stengel, Christian Waluga, and Barbara Wohlmuth.
\newblock Performance and scalability of hierarchical hybrid multigrid solvers for stokes systems.
\newblock {\em SIAM Journal on Scientific Computing}, 37(2):C143--C168, 2015.

\bibitem{Schornbaum2016}
Florian Schornbaum and Ulrich R{\"u}de.
\newblock Massively parallel algorithms for the lattice boltzmann method on nonuniform grids.
\newblock {\em SIAM Journal on Scientific Computing}, 38(2):C96--C126, 2016.

\bibitem{Succi01}
Sauro Succi.
\newblock {\em {The lattice Boltzmann equation for fluid dynamics and beyond}}.
\newblock Oxford University Press, 2001.

\bibitem{Aidun10}
Cyrus~K. Aidun and Jonathan~R. Clausen.
\newblock {Lattice-Boltzmann} method for complex flows.
\newblock {\em Annual Review of Fluid Mechanics}, 42(1):439--472, 2010.

\bibitem{djukic2023validation}
Tijana Djukic, Marko Topalovic, and Nenad Filipovic.
\newblock {Validation of lattice Boltzmann based software for blood flow simulations in complex patient-specific arteries against traditional CFD methods}.
\newblock {\em Mathematics and Computers in Simulation}, 203:957--976, 2023.

\bibitem{Chen92}
Hudong Chen, Shiyi Chen, and William~H Matthaeus.
\newblock {Recovery of the Navier-Stokes equations using a lattice-gas Boltzmann method}.
\newblock {\em Physical review A}, 45(8):R5339, 1992.

\bibitem{Qian92}
Yuehong Qian, Dominique d'Humières, and Pierre Lallemand.
\newblock {Lattice BGK models for Navier-Stokes equation}.
\newblock {\em European Journal of Mechanics B/Fluids}, 11(1):193--202, 1992.

\bibitem{Tessarotto2008}
Marco Tessarotto, Enrico Fonda, and Massimo Tessarotto.
\newblock {The Computational Complexity of Traditional Lattice-Boltzmann Methods for Incompressible Fluids}.
\newblock In {\em AIP Conference Proceedings}, volume 1084(1), pages 470--475. American Institute of Physics, 2008.

\bibitem{tau1}
Maciej Matyka and Micha{\l} Dzikowski.
\newblock {Memory-efficient Lattice Boltzmann Method for low Reynolds number flows}.
\newblock {\em Computer Physics Communications}, 267:108044, 2021.

\bibitem{Andrade99}
Jos{\'e}~S. Andrade, Uriel M.~S. Costa, Murilo~P. Almeida, Hern{\'a}n~A. Makse, and Harry~Eugene Stanley.
\newblock Inertial effects on fluid flow through disordered porous media.
\newblock {\em Physical Review Letters}, 82:5249--5252, Jun 1999.

\bibitem{He97}
Xiaoyi He and Li-Shi Luo.
\newblock {Theory of the lattice Boltzmann method: From the Boltzmann equation to the lattice Boltzmann equation}.
\newblock {\em Physical review E}, 56(6):6811, 1997.

\bibitem{Zou97}
Qisu Zou and Xiaoyi He.
\newblock {On pressure and velocity boundary conditions for the lattice Boltzmann BGK model}.
\newblock {\em Physics of fluids}, 9(6):1591--1598, 1997.

\bibitem{Tomczak2023}
Wojciech Bacza, Tadeusz Tomczak, and Maciej Matyka.
\newblock lbmtau1.
\newblock \url{https://github.com/tadeusz-tomczak/lbmtau1}, 2023.

\bibitem{LBM_principles}
Tim Kr{\"u}ger, Halim Kusumaatmaja, Alexander Kuzmin, Orest Shardt, Gonçalo Silva, and Erlend~Magnus Viggen.
\newblock {\em {The Lattice Boltzmann Method: Principles and Practice}}.
\newblock Graduate Texts in Physics. Springer International Publishing, 2018.

\bibitem{Ghia1982}
Urmila Ghia, Karman Ghia, and C.~T. Shin.
\newblock {High-Re solutions for incompressible flow using the Navier-Stokes equations and a multigrid method}.
\newblock {\em Journal of computational physics}, 48(3):387--411, 1982.

\bibitem{AMD_PPR}
Advanced Micro Devices, Inc.
\newblock {\em Processor Programming Reference (PPR) for AMD Family 19h Model 61h, Revision B1 Processors 56713 Rev 3.05}, March 2023.

\bibitem{AMD_SOG}
Advanced Micro Devices, Inc.
\newblock {\em Software Optimization Guide for the AMD Zen4 Microarchitecture}, 2023.

\bibitem{Priour2014}
DJ~Priour~Jr.
\newblock Percolation through voids around overlapping spheres: a dynamically based finite-size scaling analysis.
\newblock {\em Physical Review E}, 89(1):012148, 2014.

\bibitem{Bauer:lbm2021}
Martin Bauer, Harald K{\"o}stler, and Ulrich R{\"u}de.
\newblock {lbmpy: Automatic code generation for efficient parallel lattice Boltzmann methods}.
\newblock {\em Journal of Computational Science}, 49:101269, 2021.

\bibitem{Latt2021Pal}
Jonas Latt, Orestis Malaspinas, Dimitrios Kontaxakis, Andrea Parmigiani, Daniel Lagrava, Federico Brogi, Mohamed~Ben Belgacem, Yann Thorimbert, Sébastien Leclaire, Sha Li, Francesco Marson, Jonathan Lemus, Christos Kotsalos, Raphaël Conradin, Christophe Coreixas, Rémy Petkantchin, Franck Raynaud, Joël Beny, and Bastien Chopard.
\newblock {Palabos: Parallel Lattice Boltzmann Solver}.
\newblock {\em Computers \& Mathematics with Applications}, 81:334--350, 2021.
\newblock Development and Application of Open-source Software for Problems with Numerical PDEs.

\bibitem{Lehman:Acc2022}
Moritz Lehmann, Mathias~J Krause, Giorgio Amati, Marcello Sega, Jens Harting, and Stephan Gekle.
\newblock {Accuracy and performance of the lattice Boltzmann method with 64-bit, 32-bit, and customized 16-bit number formats}.
\newblock {\em Physical Review E}, 106(1):015308, 2022.

\bibitem{Liu2019Sun}
Zhao Liu, XueSen Chu, Xiaojing Lv, Hongsong Meng, Shupeng Shi, Wenji Han, Jingheng Xu, Haohuan Fu, and Guangwen Yang.
\newblock {SunwayLB: Enabling Extreme-Scale Lattice Boltzmann Method Based Computing Fluid Dynamics Simulations on Sunway TaihuLight}.
\newblock In {\em 2019 IEEE International Parallel and Distributed Processing Symposium (IPDPS)}, pages 557--566, 2019.

\bibitem{Qi2016Perf}
Jiaxing Qi, Kartik Jain, Harald Klimach, and Sabine Roller.
\newblock {Performance Evaluation of the LBM Solver Musubi on Various HPC Architectures}.
\newblock In {\em International Conference on Parallel Computing}, volume~27, pages 807--816, 04 2016.

\bibitem{Wittmann:Lat2018}
Markus Wittmann, Viktor Haag, Thomas Zeiser, Harald K{\"o}stler, and Gerhard Wellein.
\newblock {Lattice Boltzmann benchmark kernels as a testbed for performance analysis}.
\newblock {\em Computers \& Fluids}, 172:582--592, 2018.

\bibitem{Williams:Roof2009}
Samuel Williams, Andrew Waterman, and David Patterson.
\newblock {Roofline: an insightful visual performance model for multicore architectures}.
\newblock {\em Communications of the ACM}, 52(4):65--76, 2009.

\end{thebibliography}

\end{document}